\documentclass[12pt]{article}

\usepackage{latexsym,cmmib57}
\usepackage{amsmath,amsthm}
\usepackage[psamsfonts]{amsfonts,eucal}
\usepackage{graphicx}
\usepackage{url}
\usepackage{chicago}

\newtheorem{theorem}{Theorem}
\newtheorem{definition}{Definition}
\newtheorem{lemma}{Lemma}

\newcommand{\condgen}[6]{{#1}#2 #5 #3 #6 #4}
\newcommand{\bbrd}[1]{\mbox{\rm{I}\kern-.1667em{#1}}}

\newcommand{\PROB}{\mathbb{P}}

\newcommand{\Probcmd}[2]{\condgen{\PROB}{\Bigl\{}{\Bigm|}{\Bigr\}}{#1}{#2}}

\newsavebox{\fmbox}
\newenvironment{fmpage}[1]
        {\begin{lrbox}{\fmbox}\begin{minipage}{#1}}
        {\end{minipage}\end{lrbox}\fbox{\usebox{\fmbox}}}

\newcounter{algocnt}
\newenvironment{algolist}[1]{%
    \begin{list}{\thealgocnt}
    {\parsep 0in\usecounter{algocnt}\setcounter{algocnt}{0}\renewcommand{\thealgocnt}{{#1}\arabic{algocnt}}
    \setlength{\rightmargin}{0in}
    \settowidth{\leftmargin}{{#1}999}\addtolength{\leftmargin}{\labelsep}}}{\end{list}}
\newcommand{\algotab}[0]{\hspace*{1.2\labelsep}}
\newcommand{\mt}[0]{\algotab\ }
\newcommand{\mtt}[0]{\algotab\ \algotab\ }
\newcommand{\mttt}[0]{\algotab\ \algotab\ \algotab\ }

\newcommand{\Algo}[1]{\textsc{#1}}

\newcommand{\hori}[2]{h_{#1}(#2)}
\newcommand{\copies}[2]{g_{#1}(#2)}
\newcommand{\extinct}[1]{D_{#1}}
\newcommand{\ehori}[2]{H_{#1}(#2)}
\newcommand{\ecopies}[2]{G_{#1}(#2)}
\newcommand{\survive}[3]{p_{#1}({#3} \vert {#2})}
\newcommand{\likelihood}[2]{L_{#1}(#2)}

\begin{document}
\begin{titlepage}
\title{A probabilistic model for gene content evolution with duplication, loss, and horizontal transfer}
\author{Mikl\'os Cs\H{u}r\"os\thanks{%
		Department of Computer Science and Operations Research,
		Universit\'e de  Montr\'eal
		C.P. 6128, succ. Centre-Ville, Montr\'eal, Qu\'ebec, Canada, H3C 3J7.
		E-mail: csuros@iro.umontreal.ca.}
\and\ Istv\'an Mikl\'os\thanks{%
		Department of Plant Taxonomy and Ecology,
		E\"otv\"os L\'or\'and University, 
		1117~Budapest, 
		P\'azm\'any P\'eter S\'et\'any~1/c,
		Hungary.
		E-mail:miklosi@ramet.elte.hu}}
\maketitle
\begin{abstract}
We introduce a Markov
model for the evolution of a gene family 
along a phylogeny.
The model includes parameters for the rates of
horizontal gene transfer, 
gene duplication, and gene loss, in addition to 
branch lengths in the phylogeny.
The likelihood for the changes in the size 
of a gene family 
across different organisms 
can be calculated in~$O(N+hM^2)$ time
and~$O(N+M^2)$ space, 
where~$N$ is the number of organisms,
$h$ is the height of the phylogeny, 
and~$M$ is the 
sum of family sizes.
We apply the model to the evolution of gene content in Preoteobacteria 
using the gene families in the COG (Clusters of Orthologous Groups) 
database.
\end{abstract}
\setcounter{page}{0}
\end{titlepage}

\section{Introduction}
At this time, 
257 microbial genomes are sequenced, about 
twice as many are soon to be completed, 
and 20 complete eukaryotic genomes are publicly available
(\url{http://www.ncbi.nlm.nih.gov/Genomes/}).
These numbers continue to grow in an exponential pace with 
advances of technology and savvy~\shortcite{Sequencing.review}.
The wealth of genome sequence data already caused
a revolution in molecular evolution methods \shortcite{WolfeLi2003,phylogenomics.review}.
A few years ago, scientific studies have necessarily focused 
on nucleotide-level differences between orthologous genes, 
mainly because of the technical and financial limitations on DNA sequence 
collection. With the increasing amount of whole genome information 
it becomes possible to analyze genome-scale differences between 
organisms, and to identify the evolutionary forces responsible for 
these changes. 
In particular, sizes of 
gene families can be compared 
with the aim of better understanding adaptive evolutionary mechanisms
and organismal phylogeny.
Several studies point to the fact that 
gene content may carry sufficient phylogenetic signal 
for the construction of evolutionary 
trees \shortcite{Fitz-Gibbon1999,Snel1999,Tekaia1999,LinGerstein2000,Clarke2002,Korbel2002,Dutilh2004,Gu2004,Huson2004,Lake2004}.
Comparative analyses of genome-wide protein domain content 
\shortcite{LinGerstein2000,Yang2005,Deeds2005}
have also provided important insights into evolution.
Gene content and similar features have been used 
to construct viral \shortcite{Montague2000,Herniou2001},
microbial \shortcite{Snel1999,Fitz-Gibbon1999,Gu2004},
and universal trees \shortcite{Tekaia1999,Simonson2005,Yang2005}.
Comparative gene content analysis is also used
to estimate
ancestral genome composition \shortcite{Snel2002,Mirkin2003}. 
The presence-absence pattern of homologs in different organisms, the so-called 
phyletic pattern \shortcite{KooninGalperin,COG.new},
provides clues about gene function \shortcite{Pellegrini1999}
and evolution of metabolic pathways \shortcite{Mirkin2003}. 

There are two principal methodological issues in the analysis of gene content.
First, one needs to decide how homologous gene are selected, and secondly,
an appropriate computational technique must be chosen to analyze the data.
A possible choice for compiling a data set is to use pairwise orthologs 
and compute a score for each pair of genomes~\shortcite{Snel1999,Tekaia1999,Korbel2002,Clarke2002,Dutilh2004}.  
The matrix of pairwise scores of shared gene content is then amenable to 
analysis by distance-based methods of phylogeny construction.
An alternative 
 is proposed in \shortcite{Lake2004}: for each pair of genomes, 
the presence or absence of homologs with respect to a third reference genome is noted, 
and the pairs of presence-absence sequences are used to calculate a pairwise distance matrix.
A third approach is to first compute families of homologous genes, and then to 
record the absence or presence of each family in the genomes \shortcite{Fitz-Gibbon1999,LinGerstein2000,Jordan2001,Wolf.clades}.
The resulting absence-presence data can be treated as a set of 0-1 sequences and further 
analyzed with traditional parsimony or distance-based methods. 
Some specialized parsimony methods have been developed for the direct purpose of analyzing gene absence-presence 
data \shortcite{Mirkin2003,Kunin2003}.  
Instead of simply noting presence-absence, the gene family sizes 
give an even richer signal for evolutionary analyses~\shortcite{Snel2002,Huson2004,Hahn2005}.
It is that latter type of data that we model here.

A number of processes shape the gene content of an organism. 
New genes may be created by duplication of an existing gene, 
horizontal transfer from a different lineage, and fusion/fission \shortcite{Snel2002}.
It has been widely debated how the extent of horizontal gene transfer (HGT)
compares to vertical inheritance 
\shortcite{Jordan2001,Snel2002,Gogarten2002,Kurland2003,Kunin2005,Ge.cobweb,Simonson2005}. 
It is clear that horizontal gene transfer plays a major role in microbial evolution \shortcite{Boucher.HGT},
but there is still need for adequate mathematical models in which that role can be measured.

This paper describes a probabilistic model for gene content evolution.
Specifically, we model the evolution of 
gene families along a phylogeny. 
The model includes gene duplication, gene loss, and horizontal transfer 
as mechanisms that determine gene family evolution. 
We also show how to compute exact likelihoods 
for gene family sizes in different organisms.
A few probabilistic models were proposed for gene content evolution, but they are 
less general than ours. Usual stochastic models
work with two parameters.
Gu and Zhang~\shortcite{Gu2004,Gu2005} 
rely on a model that includes gene loss and gene duplication but no other modes of gene genesis.
They showed how gene family sizes can be used to define additive distances in such a model. 
Interestingly enough, the data can be reduced to a three-letter alphabet for the purposes of 
distance calculations: only 0, 1 or ``many'' homologs per family need to be counted. 
The distance relies on an estimate of the rate parameters, which is obtained through likelihood optimization.
Hahn {\em et al.} \shortcite{Hahn2005}
developed an alternative likelihood-based approach for the same two-parameter model.
Huson and Steel \shortcite{Huson2004} analyzed a two-parameter model 
that accounts for gene loss and horizontal transfer but not for gene duplication.
They derived a distance 
measure based on gene family sizes using likelihood maximization arguments. 
They further showed that other distance measures based on shared gene content \shortcite{Snel1999}
have inferior accuracy in phylogeny reconstruction than either Dollo parsimony or their own distance measure. 
Karev {\em et al.} developed a rich probabilistic model of 
gene content evolution in a series of papers~\shortcite{Karev.BDIM.2002,Karev.BDIM.2003,Karev.BDIM.2004}. 
The model explains the distribution of gene family sizes 
found in different organisms. It is, however, too general for 
exact detailed calculations, and for likelihood computations in particular.

To our knowledge, no tractable  stochastic model was introduced yet that accounts for 
horizontal transfer, gene loss, and duplication.
These processes cannot be modeled by using only two parameters 
because the intensity of gene loss and duplication depend on the 
size of a gene family, but the tempo by which genes are acquired through 
horizontal transfer has a constant component.
Among other applications, a model that accounts for duplication and transfer is useful 
in analyzing the evolution of metabolic networks: 
do new paths evolve by gene duplication and adaptive selection, or
by accommodating genes with new functions via horizontal gene transfer?
This paper introduces a probabilistic model
for the evolution of homologs
on a phylogeny. 
Specifically, we model the evolution of a single gene family
along the phylogeny,
where different processes 
may add new genes to the family 
or erase members of it, and arrive at 
the family sizes observed at the terminal taxa.  
We provide an algorithm that can compute
analytically the likelihood of gene family sizes
in different organisms, given an evolutionary tree. 
The algorithm calculates the likelihood of 
family sizes in
$O(N+M^2 h)$ time where~$M$ is the total number of genes in the family, 
$N$ is the number of genomes,
and~$h$ is the height of the tree.
The tree height is at most linear in~$N$,
and on average, it is $O(\sqrt{N})$ or~$O(\log N)$
for uniform or Yule-Harding distribution of random trees.
In contrast, the 
methods of~\shortcite{Gu2004} and~\shortcite{Huson2004} 
compute distances between every pair of organisms, which takes 
quadratic time in~$N$. 
The likelihood calculations of \shortcite{Hahn2005}
take cubic time in~$M$, and involve the 
evaluation of infinite sums that are truncated 
heuristically.

The article is organized in the following manner. 
Section~\ref{sec:model} introduces
our stochastic model of gene content 
evolution, and 
describes formulas 
for computing various associated probabilities,
including likelihood.
The formulas are used in an algorithm described in
Section~\ref{sec:algo}. 
Section~\ref{sec:exp} describes 
our initial experiments in modeling gene content evolution 
in 51 Proteobacteria and 3555 gene families from the COG 
(Clusters of Orthologous Groups) database~\shortcite{COG.new}.
Section~\ref{sec:conclusion} concludes the paper.
%For economy, we 
%avoid too many details in 
%the proofs. 

\section{Mathematical model}\label{sec:model}
Let~$T$ be a phylogenetic tree over a set of species~$S$. 
The tree~$T$ is a rooted tree 
with vertex set~$V(T)$ and edge set~$E(T)$, 
in which leaves are bijectively labeled with elements of~$S$.
Every edge~$e$ has a length~$t_e>0$. 
We are interested in modeling the evolution of a gene family.
The family size changes along the edges: genes may be duplicated, lost, 
or gained from an unknown source. 
We model the evolution of {\em gene counts}
(family size) at the tree nodes: 
the gene count at every node~$u\in V(T)$ is 
a random variable~$\chi(u)$ that 
can take non-negative integer values. 
In addition to the tree with its edge lengths, 
three parameters determine the joint distribution
of the gene counts:
a {\em duplication rate} $\lambda$,
a {\em loss rate} $\mu$,
and a {\em transfer rate} $\kappa$. 
The loss rate accounts for 
all possible mechanisms of gene loss, including 
deletion and pseudogenization. 
The transfer rate accounts for 
processes of gene genesis, including HGT from another lineage 
in the same tree, or HGT from 
an unknown organism. 
%We do not consider other processes 
%affecting gene content evolution such 
%as fusion/fission events, or 
%the creation of entirely new families
%through duplication and adaptive 
%evolution. We also do not discuss 
%the question of defining gene families 
%and discovering homologs between genomes.

In our model, the evolution of the gene counts 
on a branch follows a 
linear birth-and-death process \shortcite{Feller}
parametrized by~$\lambda$, $\kappa$, and~$\mu$.  
Let $\{X(t):t\ge 0\}$ denote the continuous-time Markov process
formed by the gene counts along an edge~$uv$:
$\chi(u)=X(0)$ and $\chi(v)=X(t_{uv})$.
The transition probabilities of the process are the following:
\begin{align*}
\Probcmd{X(t+\delta t)=n+1}{X(t)=n} & = \Bigl(\kappa+n\lambda \Bigr)\delta t + o(\delta t)\\*
\Probcmd{X(t+\delta t)=n-1}{X(t)=n} & = n\mu\delta t  + o(\delta t) \\*
\Probcmd{|X(t+\delta t)-n|>1}{X(t)=n} &= o(\delta t).
\end{align*}
In other words, every existing gene produces
an offspring through duplication 
with an intensity of~$\lambda$, 
or disappears with an intensity of~$\mu$,
and new genes are acquired with an intensity of~$\kappa$, 
independently from the number of existing genes.

\begin{figure}
\centerline{\includegraphics[height=0.12\textheight]{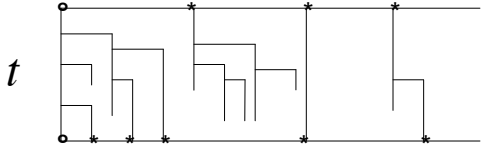}}
\caption{Galton-Watson forest 
that shows the evolution of genes 
in the same family along a tree edge.
The top 
line represents the ancestral genome with three genes; the bottom line 
represents the descendant genome, in which there are five family members.
Symbol \texttt{o} represents the source 
from which genes might be transferred horizontally, symbols $\mathtt{\star}$ represent
copies of the 
gene in the genome at the beginning and the end of the investigated time span~$t$.
Each~\texttt{o} or~$\mathtt{\star}$ in the ancestral genome 
is the root of a Galton-Watson tree.
Note that the physical order of genes in the genomes is 
immaterial: here they are simply drawn next to each other for clarity.
}\label{fig:GaltonWatson}
\end{figure}

The histories of individual genes on an edge form 
a {\em Galton-Watson} forest, see Figure~\ref{fig:GaltonWatson}.
The figure illustrates a scenario where 
the gene family increases from three to five genes.
The counts at the branch endpoints 
are the result of many duplication, transfer and loss events. 
The change involves three horizontally transferred genes, 
from among which one survives, another one does not,
and the third one 
produces two surviving paralogs. 

While it is not too difficult to 
calculate the probabilities for 
any particular gene count on a branch (see~\S\ref{ss:blocks}), 
the likelihood~$L$ of observed gene counts at the 
leaves involves an infinite number of possible gene counts
at intermediate nodes:
\begin{equation}\label{eq:likelihood.infty}
L = \sum_{\langle m_x\colon x\in V(T)\rangle} \gamma(m_{\mathsf{root}}) \prod_{xy\in E(T)} \Probcmd{\chi(y)=m_y}{\chi(x)=m_x},
\end{equation}
where $\gamma(\cdot)$ is the distribution at the root, 
and the summation over the~$\langle m_x\rangle$ vectors takes all values in agreement with the
gene counts at the leaves in the input data. 
Our main technique for computing the likelihood is
to restrict the computation to 
genes that have at least one surviving descendant 
at the leaves. In what follows we develop the 
formulas to compute the likelihood. 

\subsection{Basic transition probabilities}\label{ss:blocks}
First we analyze 
the {\em blocks} of homologs at a node 
that have a common origin.
One block is formed by 
the genes that 
trace back to a horizontal 
transfer event 
on the branch from the parent. 
Each other block is the 
set of paralogs with the same ancestor at the parent.
The homologs in Figure~\ref{fig:GaltonWatson}
form four blocks: 
a block of size three that 
comprises the descendants of the horizontally transferred genes, 
a block of size zero for the deceased parental gene, 
and two blocks of size one. 
The independent birth-and-death processes associated with the blocks 
have been thoroughly analyzed in the statistical literature. 
\begin{definition}
Define the following {\em basic transition probabilities}
for gene count evolution on a branch.
Let $\hori{t}{n}$ denote the probability that there are 
$n$ genes of foreign origin [not inherited from the parent] 
after time~$t$. 
Let $\copies{t}{n}$ denote the probability 
that a single gene has $n$ copies after time~$t$. 
\end{definition}
\begin{theorem}
The basic transition probabilities
can be written as follows. 
\begin{equation}\label{eq:hori}
\hori{t}{n} 
	= \binom{\frac\kappa\lambda+n-1}{n}
		\bigl(1-\lambda\beta(t)\bigr)^{\frac\kappa\lambda}
		\bigl(\lambda\beta(t)\bigr)^n
\end{equation}
where 
$
\beta(t) = \frac{1-e^{-(\mu-\lambda)t}}{\mu-\lambda e^{-(\mu-\lambda)t}}$,
and 
\[
\binom{\frac\kappa\lambda+n-1}{n}
	= \begin{cases}
		1 & \text{if $n=0$;}\\*
	  \frac{\bigl(\frac\kappa\lambda\bigr)\bigl(\frac\kappa\lambda+1\bigr)\dotsm \bigl(\frac\kappa\lambda+n-1\bigr)}
	{n!} & \text{if $n>0$.}
	\end{cases}
\]
%(In other words, 
%$\hori{t}{n}$ is the probability mass function for the negative binomial distribution
%with parameters~$\frac\kappa\lambda$ and~$\lambda\beta(t)$.)
Furthermore,
\begin{equation}\label{eq:copies}
\copies{t}{n} 
= \begin{cases}
	\mu\beta(t) & \text{if $n=0$;}\\*
	\bigl(1-\mu\beta(t)\bigr)\bigl(1-\lambda\beta(t)\bigr)\bigl(\lambda\beta(t)\bigr)^{n-1}
		& \text{if $n>0$.}
	\end{cases}
\end{equation}
\end{theorem}

\begin{proof}
The size of the HGT block of homologs 
follows a birth-and-death 
process with constant rate~$\kappa$ of immigration and no emigration.
The transition probabilities of \eqref{eq:hori} for 
such a process were 
analyzed by 
Karlin and McGregor~\shortcite{KarlinMcgregor}.
Blocks of paralogs evolve by a simple birth-and-death 
process: the transition probabilities of \eqref{eq:copies} 
are derived in, e.g., \shortcite{Feller}. 
\end{proof}

\subsection{Gene extinction and survival}

\begin{definition}
A {\em surviving} gene at a node~$x$ is such that 
it has at least one modern descendant at the 
leaves below~$x$.
\end{definition} 
Let $\extinct{x}$ denote the probability that a gene present at node~$x$ 
is not surviving, i.e., that it has no modern descendants. 
\begin{lemma}
The extinction probability $\extinct{x}$ can be calculated as follows.
If~$x$ is a leaf, then~$D_x=0$.
Otherwise, let $x$ be the parent of~$x_1, x_2,\dotsc, x_k$.
\begin{equation}\label{eq:extinct}
\extinct{x} =
	\prod_{j=1}^k 
	\Bigl(\mu\beta(t_j)+\bigl(1-\mu\beta(t_j)\bigr)\bigl(1-\lambda\beta(t_j)\bigr)
		\frac{\extinct{x_j}}{1-\lambda\beta(t_j)\extinct{x_j}}\Bigr)
\end{equation}
where~$t_j$ is the length of the branch leading from~$x$ to~$x_j$.
\end{lemma}
\begin{proof}
For leaves, the statement is trivial. 
When~$x$ is not a leaf, condition on the 
gene count at $x_j$:
\[
\extinct{x} = \prod_{j=1}^k
	\sum_{m=0}^{\infty} \copies{t_j}{m} \bigl(\extinct{x_j}\bigr)^m.
\]
Plugging in $\copies{t}{m}$ from Eq.~\eqref{eq:copies} 
and replacing the infinite series 
with a closed form gives~\eqref{eq:extinct}. 
\end{proof}

\subsection{Effective transition probabilities}
We introduce two new probabilities, 
denoted by
$\ehori{x}{n}$ and $\ecopies{x}{n}$.
They account for the number~$n$ of surviving genes at node~$x$,  
either acquired through horizontal transfer, or 
through duplication and loss from a single gene. 
The effective transition probabilities 
are related to $\hori{t}{n}$, and $\copies{t}{n}$, 
but account for eventual extinction 
below node~$x$. A formal definition follows. 
\begin{definition}
Let~$y$ be a non-root node and~$x$ its parent.  
Define the following 
{\em effective transition probabilities}.
Let $\ehori{y}{n}$ denote the probability that
at node~$y$ there are~$n$ surviving genes of foreign origin, 
i.e., that have no ancestor at~$x$. 
Let $\ecopies{y}{n}$ denote the probability that 
a single gene at~$x$ has~$n$ surviving copies at node~$y$. 
\end{definition}

\begin{lemma}
Let~$y$ be a non-root node, let~$x$ be its ancestor, and
let~$t$ be the length of the edge~$xy$.
The effective transition probabilities can be written as follows.
\begin{equation}\label{eq:ehori}
\ehori{y}{n}  
	= \binom{\frac\kappa\lambda+n-1}{n}
		\biggl(\frac{1-\lambda\beta(t)}{1-\extinct{y}\lambda\beta(t)}\biggr)^{\frac\kappa\lambda}
		\biggl(\frac{(1-\extinct{y})\lambda\beta(t)}{1-\extinct{y}\lambda\beta(t)}\biggr)^n
\end{equation}
\begin{subequations}\label{eq:ecopies}
\begin{align}
\ecopies{y}{0}
	& =	1-\frac{\bigl(1-\mu\beta(t)\bigr)(1-\extinct{y})}{1-\extinct{y}\lambda\beta(t)};\\*
\ecopies{y}{n}
	& =
	\frac{\bigl(1-\mu\beta(t)\bigr)\bigl(1-\lambda\beta(t)\bigr)}{\bigl(\lambda\beta(t)\bigr)(1-\extinct{y}\lambda\beta(t))}
		\biggl(\frac{(1-\extinct{y})\lambda\beta(t)}{1-\extinct{y}\lambda\beta(t)}\biggr)^{n},
		& n>0.
\end{align}
\end{subequations}
\end{lemma}
\begin{proof}
We condition on the number of 
genes at node~$y$ (whether or not they survive).
\[
\ehori{y}{n} = 
\sum_{i=0}^{\infty} \binom{n+i}{i}\hori{t}{n+i}\bigl(\extinct{y}\bigr)^i\bigl(1-\extinct{y}\bigr)^n.
\]
Using Eq.~\eqref{eq:hori} leads to an infinite series that can be 
simplified to get~\eqref{eq:ehori}. 
Similarly, write
\[
\ecopies{y}{n} = \sum_{i=0}^{\infty} \binom{n+i}{i}\copies{t}{n+i}\bigl(\extinct{y}\bigr)^i\bigl(1-\extinct{y}\bigr)^n.
\]
Taking the values of~$\copies{t}{n+i}$ from Eq.~\eqref{eq:copies} 
and simplifying the resulting infinite series yields~\eqref{eq:ecopies}. 
\end{proof}

\subsection{Number of surviving genes on a branch}
\begin{definition}
Let~$y$ be a non-root node, and let~$x$ be its ancestor.
Let $\survive{y}{n}{m}$ 
denote the {\em survival probability}
defined as the 
probability of the event that 
there are~$m$ surviving genes at node~$y$
under the condition that
there are~$n$ genes at node~$x$ (not necessarily surviving).
\end{definition}
\begin{lemma}
The survival probabilities can be computed as follows.
\begin{subequations}\label{eq:survive}
\begin{align}
\survive{y}{0}{m} & = \ehori{y}{m} 	\\*
\survive{y}{n}{0} & = \ehori{y}{0}\bigl(\ecopies{y}{0}\bigr)^n
	& 0<n \\*
\survive{y}{n}{1} & = \ecopies{y}{0}\survive{y}{n-1}{1}+\ecopies{y}{1}\survive{y}{n-1}{0}
	& 0<n \\*
\survive{y}{n}{m} & = 
	\begin{aligned}[t]
		& \alpha \survive{y}{n}{m-1}\\*
		+& \bigl(\ecopies{y}{1}-\alpha\ecopies{y}{0}\bigr)\survive{y}{n-1}{m-1}\\*
		+&\ecopies{y}{0}\survive{y}{n-1}{m}
	\end{aligned}
	& 0<n,1<m
\end{align}
\end{subequations}
where 
\begin{equation}\label{eq:alpha}
\alpha = \frac{(1-\extinct{y})\lambda\beta(t)}{1-\extinct{y}\lambda\beta(t)}.
\end{equation}
\end{lemma}
\begin{proof}
For $\survive{y}{0}{m}$ and $\survive{y}{n}{0}$, the equations are
straightforward. 
Otherwise, we condition on the surviving copies of a single gene 
at~$y$:
\begin{equation}\label{eq:survive.sum}
\survive{y}{n}{m}
	=\sum_{i=0}^m \ecopies{y}{i} \survive{y}{n-1}{m-i}.
\end{equation}
Now, 
using that $\ecopies{y}{i+1}=\alpha\ecopies{y}{i}$
whenever $i>0$,
and comparing~\eqref{eq:survive.sum} for~$\survive{y}{n}{m}$ 
and~$\survive{y}{n}{m-1}$,
we can write $\survive{y}{n}{m}$ in a recursive form as shown. 
\end{proof}

\subsection{Conditional likelihoods}\label{ss:conditional}
\begin{definition}
Let~$x$ be a node in the tree. Define 
the {\em conditional likelihood}
$\likelihood{x}{n}$ for all~$n$ 
as the probability of having the observed 
gene counts at the leaves in the subtree 
rooted at~$x$, under the condition that there 
are~$n$ surviving copies at~$x$.
\end{definition}
\begin{theorem}\label{tm:conditional}
The conditional likelihoods can be calculated as follows.
In the case when~$x$ is a leaf, $\likelihood{x}{n}=1$ 
if~$n$ is the observed gene count at~$x$, otherwise
the likelihood is~0. 
If~$x$ is not a leaf, and has children~$x_1, x_2,\dotsc, x_k$, 
then the following recursions hold.
\begin{subequations}\label{eq:likelihood}
\begin{align}
\likelihood{x}{0} &
	= \prod_{j=1}^k \sum_{m=0}^{M_j} 
		\survive{x_j}{0}{m}\likelihood{x_j}{m};\\*
\likelihood{x}{n} & 
	= (1-\extinct{x})^{-n}\biggl(\prod_{j=1}^k \sum_{m=0}^{M_j} 
		\survive{x_j}{n}{m}\likelihood{x_j}{m}\nonumber\\*
	& -\sum_{i=0}^{n-1} \binom{n}{i} (\extinct{x})^{n-i} (1-\extinct{x})^{i}\likelihood{x}{i}\biggr);
	& 0<n\le \sum_{j=1}^k M_j,
\end{align}
\end{subequations}
where~$M_j$ is the sum of gene counts at the leaves in the subtree rooted at~$x_j$.
If $n>\sum_{j=1}^k M_j$, then $\likelihood{x}{n}=0$. 
\end{theorem}
\begin{proof}
For a leaf node, or for $n>\sum_{j=1}^k M_j$, the theorem is trivial. 
Otherwise, consider the likelihood~$\ell_x(n)$ of the 
observed gene counts at the leaves in the subtree rooted at $x$, 
conditioned on the event that there are~$n$ genes present at~$x$, 
which may or may not survive. 
We write the likelihood in two ways. First, by conditioning on 
the number of surviving genes at the children,
\begin{equation}\label{eq:ell}
\ell_x(n) = \prod_{j=1}^k \sum_{m=0}^{M_j} 
		\survive{x_j}{n}{m}\likelihood{x_j}{m}.
\end{equation}
Secondly, by conditioning on the number of surviving genes at~$x$, 
\begin{equation}\label{eq:ell.sum}
\ell_x(n) = \sum_{i=0}^n \binom{n}{i} \bigl(\extinct{x}\bigr)^{n-i} \bigl(1-\extinct{x}\bigr)^i \likelihood{x}{i}.
\end{equation}
Now, rearranging the equality of the two right-hand sides 
gives the desired result. 
\end{proof}

{\sc Remark.\ \ }
Clearly, the gene counts~$M_x$ of Theorem~\ref{tm:conditional} 
are easily computed for all~$x$. If $m(x)$ is the gene count for every leaf~$x$
then
\begin{equation}\label{eq:subtree.size}
M_x = \begin{cases}
	m(x) & \text{if $x$ is a leaf;}\\*
	\sum_{j=1}^k M_{x_j} & \text{if $x_1,\dotsc, x_k$ are the children of $x$.}
\end{cases}
\end{equation}
%\begin{corollary}
%Using the likelihood~$\ell_x(n)$ from Eq.~\eqref{eq:ell}, 
%\[
%\likelihood{x}{n}
%	= 
%		\sum_{i=0}^{n} \binom{n}{i} \biggl(-\frac{\extinct{x}}{1-\extinct{x}}\biggr)^{n-i}\ell_x(i).
%\]
%\end{corollary}
%\begin{proof}
%The equation can be proven by induction using Eq.~\eqref{eq:likelihood},
%or by an inclusion-exclusion argument, inverting Eq.~\eqref{eq:ell.sum}.
%\end{proof}

\subsection{Likelihood}
It is assumed that the family size at the root is distributed according to the 
equilibrium probabilities:
\begin{equation}\label{eq:equilib}
\gamma(n) = \hori{\infty}{n} = \binom{\frac\kappa\lambda+n-1}{n} 
	\biggl(1-\frac\lambda\mu\biggr)^{\frac\kappa\lambda}
	\biggl(\frac\lambda\mu\biggr)^n.
\end{equation}
\begin{theorem}
Let~$M$ be the total number of genes at the leaves. 
The likelihood of the observed 
gene counts equals
\begin{equation}\label{eq:root}
L = \sum_{n=0}^M \likelihood{\mathsf{root}}{n} 
	\frac{\binom{\frac\kappa\lambda+n-1}{n} 
	\Bigl(1-\frac\lambda\mu\Bigr)^{\frac\kappa\lambda}
	\Bigl((1-\extinct{\mathsf{root}})\frac\lambda\mu\Bigr)^n}{\Bigl(1-\frac\lambda\mu\extinct{\mathsf{root}}\Bigr)^{\frac\kappa\lambda+n}}.	
\end{equation}
\end{theorem}
\begin{proof}
By summing the likelihoods conditioned on the surviving genes at the root,
\[
L = \sum_{n=0}^M \likelihood{\mathsf{root}}{n}\sum_{i=0}^{\infty}
	\gamma(n+i) \binom{n+i}{i} (\extinct{\mathsf{root}})^i(1-\extinct{\mathsf{root}})^n.
\]
Now, plugging in the values of $\gamma(\cdot)$ from Eq.~\eqref{eq:equilib}
and replacing the infinite series by a closed form gives the theorem's 
formula.
\end{proof}

\section{Algorithm}\label{sec:algo}
This section employs the formulas of Section~\ref{sec:model}
in a dynamic programming algorithm 
to compute the likelihood exactly.
More precisely,
the algorithm
computes the likelihood 
of gene counts at the tree leaves, given
the duplication rate~$\lambda$, 
the transfer rate~$\kappa$, and the loss rate~$\mu$. 
Algorithm \Algo{ComputeLikelihood} below proceeds by a depth-first traversal; 
the necessary variables are calculated 
from the leaves towards the root.
Let~$m(u)$ denote the gene count at every leaf~$u$.

\begin{center}
\begin{fmpage}{0.98\textwidth}
\begin{algolist}{}
\item[] \Algo{ComputeLikelihood}
\item[] \textbf{Input} $\lambda, \kappa,\mu$, $T$, gene counts $m(u)\colon \text{$u$ is a leaf of $T$}$
\item[] \textbf{Output} likelihood of the $m(\cdot)$ values
\item \textbf{for} each node $x\in V(t)$ 
	in a depth-first traversal
\item\mt\ Compute $\extinct{x}$ using Eq.~\eqref{eq:extinct}.\label{line:extinct}
\item\mt\ Compute the sum of gene counts $M_x$ by Eq.~\eqref{eq:subtree.size}.\label{line:size}
\item\mt\ \textbf{if} $x$ is not the root \textbf{then}\label{line:survive.loop}
\item\mtt\ Let $y$ be the parent of $x$.\label{line:survive.body}
\item\mtt\ \textbf{for} $n=0,\dotsc,M_y$ \textbf{do} 
\item\mttt\ \textbf{for} $m=0,\dotsc,M_x$ \textbf{do} compute $\survive{x}{n}{m}$ by Eq.~\eqref{eq:survive}.\label{line:survive}
\item\mt\ \textbf{for} $n=0,\dotsc,M_x$ \textbf{do} compute $\likelihood{x}{n}$ 
	by Eq.~\eqref{eq:likelihood}.\label{line:likelihood}
\item Compute the likelihood $L$ at the root using Eq.~\eqref{eq:root}.\label{line:root}
\item \textbf{return} $L$.
\end{algolist}
\end{fmpage}
\end{center}

Theorem~\ref{tm:algo} below analyzes the algorithm's complexity 
in terms of the topology of~$T$. In particular,  
it uses the notions of {\em height of a node}~$x$, 
defined as the number of edges on the path leading from the root to~$x$, 
{\em levels} of nodes, which are sets of nodes with the same height, and 
{\em height of the tree}, which is the maximum of the leaf heights.

\begin{theorem}\label{tm:algo}
Let~$h$ be the height of~$T$ in 
Algorithm~\Algo{ComputeLikelihood}, and~$N$
the number of its leaves,
and let~$M=M_{\mathsf{root}}$ be the sum of gene counts. 
The algorithm can be implemented 
in such a way that it uses~$O(N+M^2)$ space 
and runs in~$O(N+hM^2)$ time.
\end{theorem}
\begin{proof}
Computing~$\extinct{x}$ and~$M_{x}$ 
takes~$O(1)$ time when~$x$ is a leaf, 
or~$O(k)$ for an inner node with~$k$ children. 
There are~$O(N)$ nodes
in the tree and, thus, 
computing~$\extinct{x}$ and~$M_{x}$ 
for all~$x$ is done in~$O(N)$ time. 
The computed values are stored in~$O(N)$ space.

In order to analyze the computations in
Lines \ref{line:survive.loop}--\ref{line:likelihood}, 
we consider nodes at the same level.
Line~\ref{line:likelihood}
computes~$\likelihood{x}{n}$ for $n=0,\dotsc,M_x$
in $O(M_x^2)$ time. 
Lines~\ref{line:survive.body}--\ref{line:survive}
compute~$\survive{x}{n}{m}$ 
for $(M_x+1)(M_y+1)$ pairs of~$n,m$
values.
(Notice that~$\ehori{y}{m}$
can be computed 
in~$O(1)$ time for each~$m$ 
in the iteration over~$m$ using 
that $\ehori{y}{m} =  \alpha\frac{m+\kappa/\lambda-1}{m}\ehori{y}{m-1}$
with the~$\alpha$ of Eq.~\eqref{eq:alpha}.)
For the children~$x_1,\dotsc, x_k$ of 
the same node~$y$, 
the total time spent in 
Lines~\ref{line:survive.body}--\ref{line:survive}
is~$O(M_y^2)$.
Hence, the time spent on computing values 
for all nodes at the same level~$k$
is
\[
O\Bigl(\sum_{\text{all $y$ at level $k-1$}} M_y^2+
	\sum_{\text{all $x$ at level $k$}} M_x^2\Bigr).
\]
Clearly, $\sum_x M_x\le M$ 
if the summation goes over~$x$ for which 
their subtrees do not overlap,
such as nodes at the same level.
Now, $\sum_x M_x^2\le (\sum_x M_x)^2 \le M^2$,
and, thus, $O(M^2)$ time is spent on each level. 
Therefore, the total time 
spent in the loop of 
Line~\ref{line:survive.loop}
is~$\cdot O(hM^2)$.
Line \ref{line:root} takes~$O(M)$ time. 

In order to obtain the space complexity result, notice 
that at the end of the loop 
in Line~\ref{line:likelihood} the
computed variables for the children of~$x$
are not needed anymore. 
Therefore, the nodes 
for which~$\survive{x}{\cdot}{\cdot}$
is needed are such that their subtrees do not
overlap. 
By the same type of argument as 
with time spent on a level, 
the number of variables that 
need to be kept in memory is~$O(M^2)$.
\end{proof}

\section{Gene content evolution in Proteobacteria}\label{sec:exp}
Proteobacteria form one of the most diverse
groups of Prokaryotes. 
Proteobacteria are an excellent model case for studying genome
content evolution: they include 
pathogens, endosymbionts, and free-living organisms.
Genome sizes vary tenfold within this group,
and horizontal transfer is abundant~\shortcite{Gogarten2002}.
Their phylogeny is still 
not resolved to satisfaction~\shortcite{Lerat.gamma,Boussau.alpha,Herbeck.gamma,Belda.gamma}.
We used~51 Proteobacteria in 
the first application of our likelihood method. 
Gene counts were based on 
the newer version~\shortcite{COG.new} of the COG 
database.
Each COG is a manually curated protein family of homologs.
The COGs 
are classified into~23 functional categories. 
%COGs are routinely used in annotating newly sequenced genomes. 
For each of our 51 Protebacteria, the number of genes in each COG family was established.
There are 3555 COG families that have at least one member in the organisms.
(The organisms are listed in the Appendix.)
The data set was provided to us by 
Csaba P\'al and Martin Lercher~\shortcite{PalLercher}.
%Their study indicates that horizontal transfer contributed quite extensively to 
%shaping metabolic pathways in {\em E.~coli} in the past 100 million years. 
The purpose of applying the likelihood method was 
not to carry out in-depth data analysis, but rather 
to get a first impression of our method's performance 
on realistic data. 

\begin{figure}
\centerline{\includegraphics[width=\textwidth]{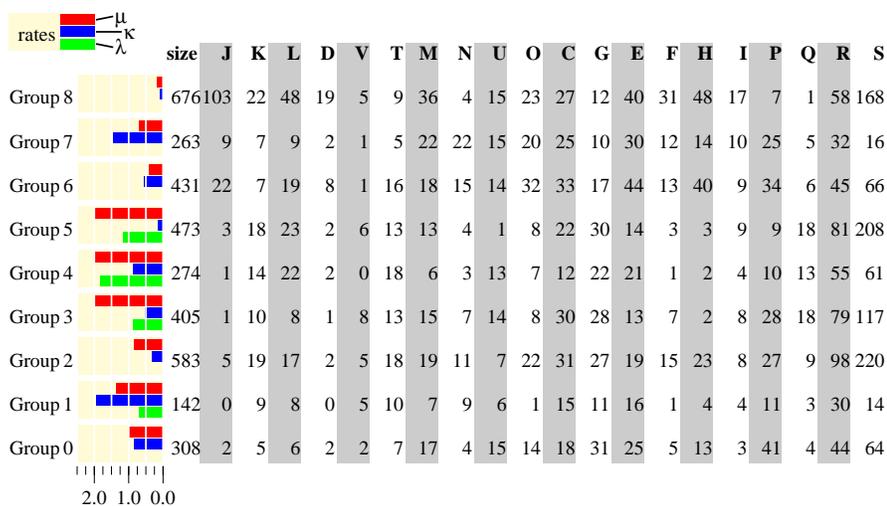}}
\caption{Rates in different groups and the distribution of COG functional categories.
The functional categories are:
J--translation, 
K--transcription, 
L--replication and repair, 
D--cell cycle control and mitosis,
V---defense mechanisms,
T--signal transduction,
M--cell wall/membrane/envelope biogenesis,
N--cell motility, 
U--intracellular trafficking and secretion, 
O--posttranslational modification, protein turnover and chaperones, 
C--energy production and conversion, 
G--carbohydrate transport and metabolism, 
E--amino acid transport and metabolism, 
F--nucleotide transport and metabolism, 
H--coenzyme transport and metabolism, 
I--lipid transport and metabolism, 
P--inorganic ion transport and metabolism, 
Q--secondary metabolites biosynthesis, transport and catabolism, 
R--general function prediction only, 
S--function unknown. 
The ``size'' columns gives 
the number of COGs in each rate 
group. (The numbers in one row do not always 
add up to the value in the ``size'' column
because some COGs have more than one 
functional assignment.)
}\label{fig:clusters}
\end{figure}

First we optimized the 
branch lengths and the~$\lambda, \kappa$ parameters 
while keeping $\mu=1.0$ to fix the scaling. 
In a second pass, we clustered the COG families 
with different rates in different groups.
The groups were established in several iterations 
of Expectation Maximization: in an E-step, each family was assigned to 
the best group (the one whose rates give the highest likelihood),
in an M-step, rates were optimized within each group separately to maximize the likelihood
of the COG gene counts within the group's families. 
Figure~\ref{fig:clusters} shows 
the rates in different groups (Groups~0--8), as well as the distribution 
of COG functional classes across clusters. 
The picture shows that 
various rate groups are needed to describe the evolution of the families. 
While the results and the methodology 
still need a thorough critical assessment, some interesting patterns already emerge.
About 19\% of the families are very stable (Group~8), this includes the large majority 
of genes involved in translation (category J) 
such as tRNA synthetases and ribosomal proteins, and cell cycle control (category D). 
About one in nine families 
fall into the groups with large horizontal transfer rates (Groups 1 and 7),
while one in three families are in groups with very low transfer rates. 
There are many categories where duplication plays only a minor role. 
For instance, the evolution of cell motility (category N), 
and various metabolic functions (F,H,I)
seem to be shaped mainly by 
horizontal transfer and loss. 

\section{Conclusion}\label{sec:conclusion}
We presented the first 
three-parameter model 
of gene content evolution,
along with a fast algorithm for 
computing likelihoods. 
We implemented 
parameter optimization 
and a gene family clustering method
and carried out a pilot experiment 
using COG family sizes in 51 Proteobacteria. 

We modeled gene family evolution by a birth-and-death process. 
It was shown that birth-and-death processes (as opposed to concerted evolution) 
appropriately represent 
family evolution in at least some gene families~\shortcite{Nei1997,Michelmore1998,Piontkovska2002}.
In addition, birth-and-death processes of various complexity explain the observed power-law behavior of 
gene family sizes~\shortcite{Karev.BDIM.2002,Karev.BDIM.2003,Karev.BDIM.2004,Reed2004}.
In order to develop a truly realistic likelihood model, 
rate variation must be permitted across lineages and families. 
Our formulas can be readily adapted to 
branch-dependent 
rates. The challenge lies rather in the parametrization: 
introducing four parameters (three rates and branch length)
for every tree edge and every family will lead to overfitting.
A possible solution is to work with 
parameters that depend only on the gene family 
and parameters that depend only on the branch. 
These two sets of parameters are combined for each branch and family 
to infer branch-specific rates.
We are now working on developing 
adequate rate-variation models, in a 
clustering-based approach as we did in Section~\ref{sec:exp}, and by 
imposing rate distribution functions. 
In another line of extension, we are investigating 
the coupling of the model with sequence evolution models to enable 
a finer modeling of homologies than simple counts. 
By scoring the similarity of genes within the same family, one can 
arrive to a finer likelihood model of gene content evolution.

It is interesting to point out that while the mathematical model 
assigns a non-zero probability to the case when 
the gene family has no members at any of the leaves, a 
family with no extant genes 
is not included usually in the data. 
Consequently, likelihood methods tend to underestimate 
the extent of gene losses. The situation is similar to 
what is encountered in likelihood models of intron evolution 
and a possible remedy is discussed in~\shortcite{Csuros.introns}.

This paper focuses on the core algorithmic 
problems of likelihood computations in 
a biologically realistic model of gene content evolution. 
The presented likelihood algorithm 
can be utilized in a number of contexts. 
The computations can be used in 
parameter optimization for estimating duplication, loss, and transfer rates 
in different gene families. 
By comparing the maximum likelihood values 
achieved with different evolutionary tree topologies, 
organismal phylogeny can be derived based on gene content. 
``Unusual'' branches with excess transfer, loss, etc., can be identified by 
examining the likelihoods, adapting an idea of \shortcite{Hahn2005}. 
The conditional likelihoods of~\S\ref{ss:conditional}
can be used in likelihood-based computations of ancestral gene content,
similarly to standard methods employed in case of molecular sequences \shortcite{Pupko.ancestral}. 
The likelihood computation enables also 
the sampling of different trees in a Bayesian Markov Chain Monte Carlo approach.
We believe that our approach to computing exact likelihoods efficiently in the three-parameter model
will find many applications in comparative gene content analysis. 

\clearpage

%\bibliographystyle{chicago}
%\bibliography{journals,genecontent}

\clearpage
\appendix

\section*{Appendix: organisms in the data set}

The picture below shows the organisms and the phylogeny 
in the experiments of Section~\ref{sec:exp}. Branch lengths 
are already optimized to maximize the likelihood.
Notice that branch lengths are not easy to interpret: 
scaling is defined in such a way that the
rate~$\mu=1$ in Group~0, a modestly dynamic group (cf.\ Fig.~\ref{fig:clusters}). 

\centerline{\includegraphics[height=0.8\textheight]{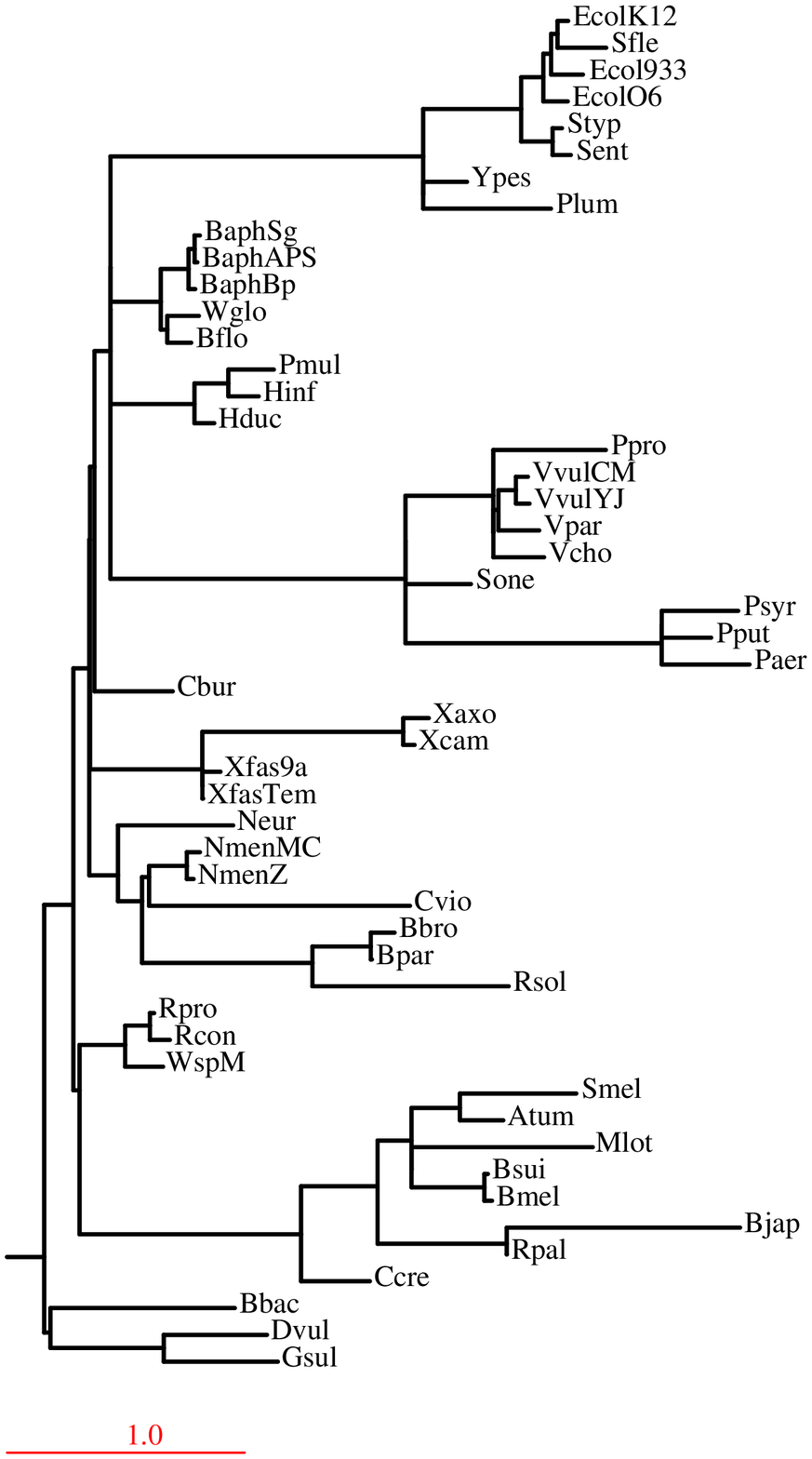}}

Abbreviations: 
EcolK12	--{\em Escherichia coli}~K12,
Sfle	--{\em Shigella flexneri}~2a str. 2457T,
Ecol933	--{\em Escherichia coli}~O157:H7 str. EDL933,
EcolO6	--{\em Escherichia coli}~O6,
Styp	--{\em Salmonella typhimurium}~LT2,
Sent	--{\em Salmonella enterica} subsp. {\em enterica serovar Typhi}~str. CT18,
Ypes	--{\em Yersinia pestis biovar Medievalis}~str. 91001,
Plum	--{\em Photorhabdus luminescens} subsp. {\em laumondii}~TTO1,
BaphSg	--{\em Buchnera aphidicola}~str. Sg,
BaphAPS	--{\em Buchnera aphidicola}~str. APS,
BaphBp	--{\em Buchnera aphidicola}~str. Bp,
Wglo	--{\em Wigglesworthia glossinidia} endosymbiont of {\em Glossina brevipalpis},
Bflo	--{\em [Candidatus] Blochmannia floridanus},
Pmul	--{\em Pasteurella multocida} subsp. {\em multocida}~str. Pm70,
Hinf	--{\em Haemophilus influenzae}~Rd KW20,
Hduc	--{\em Haemophilus ducreyi}~35000HP,
Ppro	--{\em Photobacterium profundum}~SS9,
VvulCM	--{\em Vibrio vulnificus}~CMCP6,
VvulYJ	--{\em Vibrio vulnificus}~YJ016,
Vpar	--{\em Vibrio parahaemolyticus}~RIMD 2210633,
Vcho	--{\em Vibrio cholerae} O1 {\em biovar eltor}~str. N16961,
Sone	--{\em Shewanella oneidensis}~MR-1,
Psyr	--{\em Pseudomonas syringae}~pv. tomato str. DC3000,
Pput	--{\em Pseudomonas putida}~KT2440,
Paer	--{\em Pseudomonas aeruginosa}~PAO1,
Cbur	--{\em Coxiella burnetii}~RSA 493,
Xaxo	--{\em Xanthomonas axonopodis}~pv. {\em citri} str. 306,
Xcam	--{\em Xanthomonas campestris}~pv. {\em campestris} str. ATCC 33913,
Xfas9a	--{\em Xylella fastidiosa}~9a5c,
XfasTem	--{\em Xylella fastidiosa}~Temecula1,
Neur	--{\em Nitrosomonas europaea}~ATCC 19718,
NmenMC	--{\em Neisseria meningitidis}~MC58,
NmenZ	--{\em Neisseria meningitidis}~Z2491,
Cvio	--{\em Chromobacterium violaceum}~ATCC 12472,
Bbro	--{\em Bordetella bronchiseptica}~RB50,
Bpar	--{\em Bordetella parapertussis}~12822,
Rsol	--{\em Ralstonia solanacearum}~GMI1000,
Rpro	--{\em Rickettsia prowazekii}~str. Madrid E,
Rcon	--{\em Rickettsia conorii}~str. Malish 7,
WspM	--{\em Wolbachia} endosymbiont of {\em Drosophila melanogaster},
Smel	--{\em Sinorhizobium meliloti}~1021,
Atum	--{\em Agrobacterium tumefaciens}~str. C58,
Mlot	--{\em Mesorhizobium loti}~MAFF303099,
Bsui	--{\em Brucella suis}~1330,
Bmel	--{\em Brucella melitensis}~16M,
Bjap	--{\em Bradyrhizobium japonicum}~USDA 110,
Rpal	--{\em Rhodopseudomonas palustris}~CGA009,
Ccre	--{\em Caulobacter crescentus}~CB15,
Bbac	--{\em Bdellovibrio bacteriovorus}~HD100,
Dvul	--{\em Desulfovibrio vulgaris} subsp. {\em vulgaris}~str. Hildenborough,
Gsul	--{\em Geobacter sulfurreducens}~PCA.

\end{document}